\documentclass[twocolumn]{elsart}

\usepackage{graphics}
\usepackage{amssymb}

\begin{document}

\begin{frontmatter}

% Title, authors and addresses

% use the thanksref command within \title, \author or \address for footnotes;
% use the corauthref command within \author for corresponding author footnotes;
% use the ead command for the email address,
% and the form \ead[url] for the home page:
% \title{Title\thanksref{label1}}
% \thanks[label1]{}
% \author{Name\corauthref{cor1}\thanksref{label2}}
% \ead{email address}
% \ead[url]{home page}
% \thanks[label2]{}
% \corauth[cor1]{}
% \address{Address\thanksref{label3}}
% \thanks[label3]{}

\title{LHCb Level-0 Trigger Detectors}

% use optional labels to link authors explicitly to addresses:
% \author[label1,label2]{}
% \address[label1]{}
% \address[label2]{}

\author{A.~Sarti\corauthref{cor}}

\corauth[cor]{Corresponding author. Address: asarti@lnf.infn.it}

\address{INFN-Laboratori Nazionali di Frascati, Via E.Fermi, 40, 44100, Frascati, Italy}

\begin{abstract}
% Text of abstract
The calorimeter and muon systems are essential components 
to provide a trigger for the LHCb experiment.  The calorimeter system
comprises a scintillating pad detector and pre-shower, followed 
by electromagnetic and hadronic calorimeters. The 
calorimeter system allows photons, electrons and hadrons to be 
identified, and their energy to be measured. The muon system consists 
of five measuring stations equipped with Multi-Wire Proportional
Chambers (MWPCs) and triple-Gas Electron Multiplier (GEM) 
detectors, separated by iron 
filters. It allows the muons identification and transverse momentum 
measurement. The status of the two systems and their 
expected performance is presented.
\end{abstract}

\end{frontmatter}

% main text
\section{Introduction}
\label{sec:intro}

The LHCb experiment is dedicated 
to the study of the decays of beauty hadrons
produced at the LHC. Precision measurements
of CP violation and rare decays in the 
$B$ meson systems, which are the main LHCb goals, can be achieved
only with a very well designed and efficient trigger\cite{bib:trigger}.\\
A key role in the trigger is played by the Level-0 (L0)
hardware step that reduces the event rate from 40MHz to
1MHz using the input from the VELO\cite{bib:VELO} detector, the calorimeter
system\cite{bib:calotdr} and the muon system \cite{bib:muontdr}.
Design performances have been optimized 
in order to allow a fast reconstruction of high 
transverse energy ($E_T$, CALO system)
or high transverse momentum ($P_T$, MUON system) candidates
needed by the L0 decision unit (L0DU).

\section{Calorimeter system}
\label{sec:Calo}

The CALO system\cite{bib:calotdr} is made of a Scintillating
Pad Detector (SPD), followed by a Pre-Shower (PS)
in front of the Electromagnetic CALorimeter (ECAL) and
the Hadronic CALorimeter (HCAL).\\
The active elements
are scintillating tiles, read out via wave-length shifting fibres 
to photo-multipliers.  The scintillator is interleaved with lead for 
the electromagnetic calorimeter in a Shashlik construction, while for 
the hadronic calorimeter it is interleaved with steel tiles.\\
All these subdetectors are characterised by a pseudo-projective 
geometry achieved using a variable detector module granularity.

\subsection{Scintillating pad detector and pre-shower}
\label{sec:SPD-PS}

The SPD and PS detectors are used to distinguish
electrons from pions and photons (SPD), photons from
Minimum Ionizing Particles and to veto busy events
with a very high charged multiplicity (SPD).\\
The SPD/PS system is made from a lead converter plate 
(14mm thick) that is  sandwiched between two layers of 
scintillator pads (15mm thick).
The light collected is read from a Multi-Anode
PhotoMulTiplier (MAPMT) and sent to a Front-End card
(VFE).

\subsection{Electromagnetic calorimeter}
\label{sec:ECAL}
%==============================================
\begin{figure}[htbp]
\begin{center}
\resizebox{6.5cm}{!}{%
\includegraphics{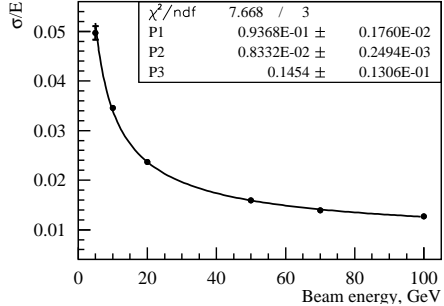}
}
\caption{ECAL module relative energy resolution, measured with
a 50GeV electron beam.}
\label{fig:energy-reso}
\end{center}
\end{figure}
%==============================================
The electromagnetic calorimeter is used to measure
the electron, photon and pion $E_T$. It provides
particle identification (PID) and reconstruction
information for the particles used in offline analysis.\\
It consists of nearly 3k Shashlik modules, for
a total of 6k detector cells,
that are read using WaveLenght Shifting (WLS) fibers from
KURARAY. Each module is positioned on the ECAL wall 
with $\pm$0.5mm ($x,y$) and $\pm$2mm ($z$) precision.\\ 
In order to achieve the design relative energy
resolution ($\sigma(E)/E = 10\% / \sqrt{E} \oplus 1.5\%$, 
with $E$ in GeV), a sampling structure of 2mm lead sheets
interspersed with 4mm thick scintillator plates has been used:
test beam results, using 50GeV electrons, are shown in
Fig.\ref{fig:energy-reso} (the achieved $\sigma(E)/E$ is 
$(9.4\pm0.2)\% / \sqrt{E} \oplus (0.83\pm0.02)\% \oplus (0.145\pm0.013)/E$, 
with $E$ in GeV).\\
Each module light yield is measured to be nearly 
3000 ph.e./GeV.
Cell dimensions are varying in the range
4$\times$4cm$^2$ to 12$\times$12cm$^2$.
The ECAL depth accounts for 25 electromagnetic radiation
lengths ($X_0$) and 1.1 hadronic interaction lengths
($\lambda$).

\subsection{Hadronic calorimeter}
\label{sec:HCAL}

The hadronic calorimeter is an iron-scintillator
tile calorimeter, composed of 52 modules
(1468 detector cells) with
variable granularity: the cell dimensions
vary from 13$\times$13cm$^2$ to 26$\times$26cm$^2$.\\
It is used to measure the $E_T$ of hadrons, and to provide
them to the offline analysis.
The HCAL depth accounts for 5.6 $\lambda$:
each module is read using WLS fibers and PMTs.\\
HCAL modules have been tested with a 30GeV electron
beam: the signal pulse is well contained in the 
25ns bunch-crossing window. The energy resolution
achieved matches the design value 
($\sigma(E)/E = 80\% / \sqrt{E} \oplus 10\%$, 
with $E$ in GeV) while
the achieved tile-to-tile spread is less than 5\%.\\
The HCAL wall has been assembled achieving 
a module-positioning precision of the order of
0.5mm ($x,y$ plane and $y$ coordinate) and 1.5mm ($y,z$ lateral
plane). 

\section{Muon system}
\label{sec:Muon}

The LHCb muon system \cite{bib:muontdr}
is composed of five tracking stations, each subdivided in 
four concentric regions, which comprise 1368 MWPCs
and 24 triple GEM detectors.
The muon detector is required to have a high detection efficiency and
a good spatial and time resolution.\\
The geometry is projective: the layout 
has been optimized by choosing an $x,y$ granularity 
that added a contribution to the $P_T$ resolution 
nearly equal to that from the multiple
scattering.
The final layout is composed of 20 different pad sizes
(from 6.3$\times$31mm to 250$\times$310mm) resulting in 120k
logical channels.\\

\subsection{Triple GEM detectors}
\label{sec:GEM}

Triple GEM detectors\cite{bib:GEM} are going to be used only in the 
innermost region of the first station,
where the particle rate is expected to be the highest.\\
Those chambers are characterised by a high
rate capability, a high station efficiency 
(greater than 96\% in 20ns time window)
and a low cluster size (less than 1.2).
The achieved time resolution is 3ns when using 
an Ar/CO$_2$/CF$_4$ gas mixture of 45\%/15\%/40\%.
%a gas mixture composed by Ar/CO$_2$/CF$_4$ (45/15/40).

\subsection{MWPC}
\label{sec:MWPC}

%
%==============================================
\begin{figure}[!t]
\begin{center}
\resizebox{6.5cm}{!}{%
\includegraphics{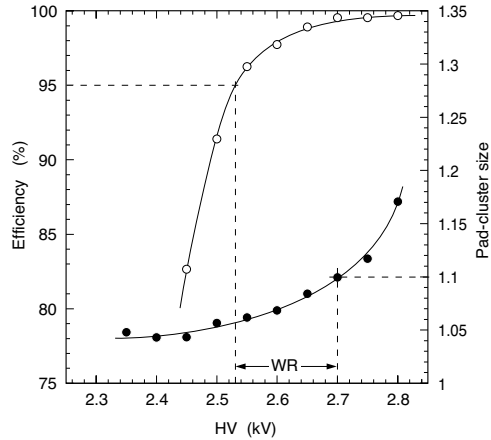}
}
\caption{Efficiency (left scale and open circles) and pad-cluster size 
(right scale and solid circles) of a double-gap
 MWPC as a function of the high voltage (HV). The working region (WR)
is shown. Curves are drawn to guide the eye.}
\label{fig:plateau}
\end{center}
\end{figure}
%==============================================
%
The MWPCs  used in the muon system
are composed of double gas gaps that
are further logically OR-ed:
this will ensure a high hit detection efficiency and enhance the
detector robustness.\\
In order to meet the performance required for triggering and for physics
analysis, each double-gap should have an
efficiency $\geq 95$\%, within a 20ns time window, 
a cluster size $\leq 1.1$, as well as good ageing properties. 
These performances were measured, using a MIP test beam, 
as a function of the high voltage (HV) applied to the wires,
using an Ar/CO$_2$/CF$_4$ gas mixture of 40\%/50\%/10\%.
The results reported in Fig.\ref{fig:plateau}
show that the above conditions are satisfied for 
$2530 \leq \mathrm{HV} \leq 2700$V.\\
MWPCs are read out with a custom radiation hard chip (named
CARIOCA) 
that works as an amplifier, shaper and discriminator
characterised by a 10ns peak time. 
Chambers installation is currently starting: the expected
positioning precision is $\pm$1mm.

\section{L0 trigger}

The L0 hardware trigger\cite{bib:trigger} is fully synchronous, 
it has a fixed 
latency (4$\mu$s) and reduces the 40MHz interaction rate to
1MHz.  The L0 Decision Unit (L0DU) takes as input: (from the CALO)
the highest $E_T$ candidate for each type (e, $\gamma$, $\pi^0$),
the total measured energy, the charged multiplicity
and (from the MUON) the two highest $P_T$ candidates per quadrant.
An efficiency of 30-50\% is achieved for the hadronic channels
(accounting for about 700kHz bandwidth) while
the muon channels are selected with 90\% efficiency.

\section{Conclusion}

The LHCb CALO and MUON systems 
are currently under installation and are going
to be ready for the data taking start in July 2007.
The calorimeter construction is now almost completed, and more than 
half of the required muon chambers have been produced.  
The muon filters are already in place while the support
wall is being built. 
Test beam data showed that the
design values have been achieved: detectors timing and 
resolution performances are very good.
The L0 hardware trigger design has been finalized:
the obtained efficiencies for hadronic, muonic and
radiative channels are expected to be 30--50\%, 90\% and 70\% 
respectively.

\end{document}